\documentclass[runningheads,envcountsame]{llncs}

  \usepackage[utf8]{inputenc}
  \usepackage{microtype}
  \usepackage[greek,british]{babel}
  \babeltags{gr=greek}
  \usepackage[hidelinks]{hyperref}

  \usepackage{amsmath}
  \usepackage{amssymb}
  \usepackage{mathtools}
  \usepackage{scalerel}
  \usepackage{tikz}
  \usepackage{newfloat}
  \usepackage{relsize}
  \usepackage{fancyhdr}

  \makeatletter
    \newsavebox{\slant@box}
    \newcommand{\slantbox}[2][.5]{\mbox{\sbox{\slant@box}{#2}%
            \hskip\wd\slant@box\pdfsave
            \pdfsetmatrix{1 0 #1 1}%
            \llap{\usebox{\slant@box}}%
            \pdfrestore}}
  \makeatother
  \newcommand{\slantedAleph}{\kern-0.02ex\slantbox[.3]{$\aleph$}\kern0.02ex}

  \newcommand{\ruleName}[1]{\textsc{\smaller({#1})}}
  \newcommand{\bnfPrim}[1]{\ensuremath{\textsc{\smaller{#1}}}}
  \newcommand{\atom}[1]{\ensuremath{\textsc{{#1}}}}
  \newcommand{\haltSym}{\textbf{!}}
  \newcommand{\unit}{(\hspace{-0.3ex})}
  \newcommand{\unif}[2][1]{\mathrel{\overset{#2}{\scalebox{#1}[1]{$\sim$}}}}
  \newcommand{\ruleto}[1]{\mathrel{\overset{\kern-0.5ex #1}{\to}}}
  \newcommand{\comment}[1]{\textit{\smaller-{}- #1}}

  \DeclareFloatingEnvironment[within=none]{listing}

\begin{document}

\title{The $\aleph$-Calculus}
\subtitle{A declarative model of reversible programming}
\author{Hannah Earley\orcidID{0000-0002-6628-2130}}
\authorrunning{H.\ Earley}
\institute{Department of Applied Mathematics, University of Cambridge, UK
  \email{h.earley@damtp.cam.ac.uk}}
\maketitle

\fancypagestyle{proc}{\fancyhf{}\renewcommand{\headrulewidth}{0pt}%
  \fancyfoot[L]{Accepted for publication in the proceedings of the 14th Conference on Reversible Computation}}
\thispagestyle{proc}

\begin{abstract}

  A novel model of reversible computing, the $\aleph$-calculus\footnotemark, is introduced.
  It is declarative, reversible-Turing complete, and has a local term-rewriting semantics.
  Unlike previously demonstrated reversible term-rewriting systems, it does not require the accumulation of history data.
  Terms in the $\aleph$-calculus, in combination with the program definitions, encapsulate all program state.
  An interpreter was also written.


  \footnotetext{%
    The name of the calculus is inspired by the Greek meaning `not forgotten', {\gr ἀλήθεια}.}
  
\end{abstract}

\section{Introduction}
\label{sec:intro}

Reversible computing is a response to Szilard's and Landauer's observations~\cite{landauer-limit,szilard-engine} that the `erasure' of information, as is common in conventional computing, leads to a fundamental thermodynamic cost in the form of an entropy increase.
To avoid this, one should ensure that the computational state transitions are \emph{injective}, i.e.\ every valid%
  \footnote{No such constraint need be applied to invalid computational states~\cite{frank-rc-generalised}.}
computational state has at most one valid predecessor.
This logical reversibility thus circumvents the Landauer-Szilard limit by avoiding the need to erase information via logical reversibility of its operation.

We introduce a novel model of reversible computing, the $\aleph$-calculus.
It is declarative, in that programs describe the logic of the computation whilst the control flow is left implicit.
Irreversible declarative languages include \texttt{Prolog}, but we believe the $\aleph$-calculus is the first declarative model of reversible computing.
Its semantics are that of a Term-Rewriting System (TRS):
  A given computation is represented by a term, and this is reversibly transformed by a transition rule to complete the computation.
Reversible TRSs have been studied before.
For example, Abramsky~\cite{abramsky-struct} introduces a general approach to modeling reversible TRSs, although he then applies it to reversibly simulating the irreversible $\lambda$-calculus.
To reversibly simulate an irreversible system, some computational history must be recorded.
This is implicit in Abramsky's treatment, but explicit in the approaches of Di Pierro et al.~\cite{rcl} and Nishida et al.~\cite{rev-term-rewriting}.
We believe ours is the first reversible-Turing complete TRS that doesn't require such recording.

In Section~\ref{sec:ex} we introduce the $\aleph$-calculus by example.
In Sections~\ref{sec:dfn} and~\ref{sec:sem} we formalize the model's definition and its semantics.
We conclude with Section~\ref{sec:concl} and briefly discuss some language extensions.
In the interest of space, proofs of theorems such as non-ambiguity and reversible-Turing completeness are deferred to an accompanying longer report.

\section{Examples}
\label{sec:ex}

\subsubsection{Recursion: Addition \& Subtraction}

The inductive definition of Peano addition, $a+\atom{Z}=a$ and $a+\atom{S} b=\atom{S}(a+b)$ where a natural number is either $\atom{Z}\equiv 0$ or the successor (\atom{S}) of another natural number, can be readily reversibilized.
Reversible addition must return both the sum, and additional information to determine what both addends were.
Here, we will keep the second addend: $a+\atom{Z}=(a,\atom{Z})$ and $a+b=(c,b) \implies a+\atom{S} b=(\atom{S} c, \atom{S} b)$.
In the $\aleph$-calculus, this is:
  \begin{align*}
    \haltSym~& {+}~{a}~{b}~{\unit}{;} \quad \haltSym~{\unit}~{c}~{b}~{+}{;} \\
    & {+}~{a}~{\atom{Z}}~{\unit} = {\unit}~{a}~{\atom{Z}}~{+}{;} && \ruleName{add--base}  \\
    & {+}~{a}~(\atom{S} b)~{\unit} = {\unit}~(\atom{S} c)~(\atom{S} b)~{+}{:} && \ruleName{add--step}  \\
    &\qquad  {+}~{a}~{b}~{\unit} = {\unit}~{c}~{b}~{+}{.} && \ruleName{add--step--sub} 
  \end{align*}
The first line says that terms of the form ${+}~a~b~\unit$ and $\unit~c~b~{+}$ are halting states (either initial or final), where $a,b,c$ are variable terms.
This is stated via the syntax $\haltSym$.
Note the $\unit$ term, pronounced `unit'; this is an empty set of parentheses that is used by convention for disambiguation between definitions.
The second line, definition \ruleName{add--base}, implements the base rule that $a+\atom{Z}=(a,\atom{Z})$.
As this is a Term-Rewriting System (TRS), we have a term on both sides and so the expression is more symmetric.
As the term encapsulates all information about program state, we need a \emph{witness} to the fact that an addition (rather than a multiplication or something else) was performed; here this is given by reusing the $+$ symbol (identifier) on the right hand side.
The next definition, \ruleName{add--step}, implements the inductive step and has as a \emph{sub-rule} \ruleName{add--step--sub}, which performs the recursion.
The above can perhaps best be understood through the below evaluation trace of $3+2=(5,2)$:
%
% BEWARE, MAGIC NUMBERS LIE WITHIN
\begingroup%
\small%
\newcommand{\subLeft}[1][1]{\smash{\raisebox{-0.65em}{\scaleto{\begin{tikzpicture}%
    \draw[<->] (0,0) to [out=270,in=90, looseness=1] (-#1,-0.5);%
  \end{tikzpicture}}{1.78em}}}\hspace{5.6em}}%
\newcommand{\subRight}[1][1]{\hspace{9.6em}\smash{\raisebox{-0.65em}{\scaleto{\begin{tikzpicture}%
    \draw[<->] (#1,-0.5) to [out=90,in=270, looseness=1] (0,0);%
  \end{tikzpicture}}{1.78em}}}}%
\newcommand{\bindings}[1]{\{#1\}}%
\newcommand{\bindingsSub}[1]{\smash{\underline{\bindings{#1}}}}%
\newcommand{\subterm}[1]{\smash{\overline{#1}}}%
\newcommand{\midsp}{\phantom{\bindings{a\mapsto3}}}%
\begin{align*}
  \haltSym\quad{+}~{3}~{2}~{\unit} \leftrightsquigarrow \bindingsSub{a\mapsto3, b\mapsto1} &\midsp
    \bindingsSub{c\mapsto4, b\mapsto1} \leftrightsquigarrow {\unit}~{5}~{2}~{+}\quad\haltSym &&
    \ruleName{add--step} \\
  \subLeft & \hspace{-0.3em}\subRight && \ruleName{add--step--sub} \\
  \subterm{{+}~{3}~{1}~{\unit}} \leftrightsquigarrow \bindingsSub{a\mapsto3, b\mapsto0} &\midsp
    \bindingsSub{c\mapsto3, b\mapsto0} \leftrightsquigarrow \subterm{{\unit}~{4}~{1}~{+}} &&
    \ruleName{add--step} \\
  \subLeft[0.25]\hspace{-2.8em} & \hspace{-3.3em}\subRight[0.25] &&
    \ruleName{add--step--sub} \\
  \subterm{{+}~{3}~{\atom{Z}}~{\unit}} \leftrightsquigarrow{} &\bindings{a\mapsto3} \leftrightsquigarrow
    \subterm{{\unit}~{3}~{\atom{Z}}~{+}} &&
    \ruleName{add--base}
\end{align*}%
\endgroup%
Squiggly arrows represent matching against the patterns in the definitions, and solid arrows refer to instantiation/consumption of `sub-terms'.

\subsubsection{Iteration: Squaring \& Square-Rooting}

A more involved example is given by reversible squaring and square-rooting.
Reversible squaring may be implemented using addition as a sub-routine via the fact $m^2=\sum_{k=0}^{m-1} (k+k+1)$.
By doing the sum in reverse (i.e.\ starting with $k=m-1$ and decrementing it towards $0$), $m$ is consumed whilst $n=m^2$ is generated.
This also shows how iteration/looping can be implemented in the $\aleph$-calculus.
The definition of squaring (\atom{Sq}) is given below:
  \begin{align*}
    \haltSym~& {\atom{Sq}}~{m}~{\unit}{;} \quad \haltSym~{\unit}~{n}~{\atom{Sq}}{;} \\
    & {\atom{Sq}}~{m}~{\unit} = {\atom{Sq}}~{\atom{Z}}~{m}~{\atom{Sq}}{;} && \ruleName{sq--begin} \\
    & {\atom{Sq}}~{s}~{(\atom{S} k)}~{\atom{Sq}} = {\atom{Sq}}~{(\atom{S} s'')}~{k}~{\atom{Sq}}{:} && \ruleName{sq--step} \\
    &\qquad {+}~{s}~{k}~{\unit} = {\unit}~{s'}~{k}~{+}. &&\comment{$\mathrlap{s'}\phantom{s''}\leftarrow \mathrlap{s}\phantom{s'}+k$} \\
    &\qquad {+}~{s'}~{k}~{\unit} = {\unit}~{s''}~{k}~{+}. &&\comment{$s''\leftarrow s'+k$} \\
    & {\atom{Sq}}~{n}~{\atom{Z}}~{\atom{Sq}} = {\unit}~{n}~{\atom{Sq}}{;} && \ruleName{sq--end}
  \end{align*}
In a reversible loop, you have a reverse conditional branch for entering/continuing the loop and a forward conditional branch for continuing/exiting the loop.
These are implemented by rules \ruleName{sq--begin,--end}.
Meanwhile rule \ruleName{sq--step} performs the actual additions of the sum.
The reader may notice that \atom{Sq} appears twice in \ruleName{sq--step}; this is merely for symmetric aesthetics, and to distinguish it from halting terms which are typically marked with $\unit$.
An example evaluation trace of $3^2=9$ (resp.\ $\sqrt{9}=3$) is given by:
\begin{align*}
  \mathllap{\haltSym\quad}
  {\atom{Sq}}~{3}~{\unit}
  = {\atom{Sq}}~{\atom{Z}}~{3}~{\atom{Sq}}
  = {\atom{Sq}}~{5}~{2}~{\atom{Sq}}
  = {\atom{Sq}}~{8}~{1}~{\atom{Sq}}
  = {\atom{Sq}}~{9}~{\atom{Z}}~{\atom{Sq}}
  = {\unit}~{9}~{\atom{Sq}}
  \mathrlap{\quad\haltSym}
\end{align*}
There is no precondition in the forward direction (except that $m$ should be a well-formed natural number), but the reverse direction requires $n=m^2$ be a square number.
If we try to take $\sqrt{10}$,
\begin{align*}
  \mathllap{\haltSym\quad}
  {\unit}~{10}~{\atom{Sq}}
  = {\atom{Sq}}~{10}~{\atom{Z}}~{\atom{Sq}}
  = {\atom{Sq}}~{9}~{1}~{\atom{Sq}}
  = {\atom{Sq}}~{6}~{2}~{\atom{Sq}}
  = {\atom{Sq}}~{1}~{3}~{\atom{Sq}}
  \mathrlap{\quad\boldsymbol\bot}
\end{align*}
the computation stalls with $\{s''\mapsto0,k\mapsto3\}$ because $3$ cannot be subtracted from $0$.
Specifically, there is no matching rule for the sub-term ${\unit}~{\atom{Z}}~{3}~{+}$.

\section{Definition}
\label{sec:dfn}

A computation in the $\aleph$-calculus consists of a \emph{term} and a \emph{program} governing the reversible evolution of the term.
A term is a tree whose leaves are \emph{symbols}, which are drawn from some infinite set of identifiers (e.g.\ $+$, \atom{Sq}, \atom{S}, \atom{Z}, \atom{Map});
  terms are conventionally written with nested parentheses, e.g.\ ${\atom{Sq}}~(\atom{S} (\atom{S} (\atom{S} \atom{Z})))~{\unit}$.
Equivalently, a term is either a symbol or a string of terms; we call a string of terms a \emph{multiterm}.
A program is a series of \emph{definitions}.
A definition is either
  (1) \emph{halting}, e.g.\ $\haltSym~{\atom{Sq}}~{n}~{\unit}$, indicating that multiterms matching the given pattern are in a halting state (annotated with $\haltSym$); or
  (2) \emph{computational}, e.g.\ ${\atom{Sq}}~{n}~{\unit} = {\atom{Sq}}~{\atom{Z}}~{n}~{\atom{Sq}}$, indicating that multiterms matching the left-pattern may be mapped to a multiterm matching the right-pattern, or vice-versa.
Computational definitions may have sub-rules, e.g.\ ${+}~{s}~{k}~{\unit} = {\unit}~{s'}~{k}~{+}$, which need to be invoked to complete the mapping.
These are summarized in BNF notation below:
\begin{align}
  \tag*{\ruleName{pattern term}} \pi &::= \bnfPrim{sym} ~|~ \bnfPrim{var} ~|~ (\,\pi^\ast\,) \\
  \tag*{\ruleName{rule}} \rho &::= \pi^\ast = \pi^\ast \\
  \tag*{\ruleName{definition}} \delta &::= \rho : \rho\rlap.^\ast ~|~ \haltSym~{\pi^\ast}{;}
\end{align}

\subsubsection{Determinism Constraints}

Like other models of reversible computation, such as Bennett's reversible Turing Machine~\cite{bennett-tm}, constraints must be placed on which programs are accepted to ensure unambiguously deterministic reversibility.
That is, whilst all the computational definitions are locally reversible---in the sense that their specific action can be uniquely reversed up to sub-rule determinism---there may be a choice of computational rules.
Whilst the semantics (Section~\ref{sec:sem}) preclude ambiguity, it is useful to reject ambiguous programs before execution.
Ensuring determinism is essentially the same as Bennett's approach: the domains and codomains of transitions must be unambiguous.
This is slightly complicated in the $\aleph$-calculus because computational definitions are bidirectional: each definition specifies two rules, one corresponding to the `forward' direction of the rule and one to its inverse.
This means that domains and codomains of definitions are conflated and a multiterm may generally match up to two rules: the `intended' rule, and the inverse of the rule which produced it.
This would lead to ambiguity, except that the $\aleph$-calculus has a notion of \emph{computational inertia} (Definition~\ref{def:inertia}).
The consequence of computational inertia is that, although a multiterm may match up to two rules, there is always a unique choice of which rule to apply at each step for a \emph{given} direction of computation.
A necessary and sufficient condition for avoiding ambiguity is given by Theorem~\ref{thm:nonamb}, and an accompanying algorithm for static analysis of $\aleph$ programs that verifies this condition is provided in the forthcoming extended version of this paper.

\begin{definition}[Term Reduction]
  \label{def:reduction}
  In the $\slantedAleph$-calculus, an input \emph{halting} multiterm $t_0$ is `reduced' to an output halting multiterm $t_n$, where $n\in\mathbb N\cup\{\infty\}$, by a series of $n$ rules $r_{i\to i+1}$.
  Each rule $r$ has a unique inverse $r^{-1}$, which gives a trivial inverse reduction from $t_0'=t_n$ to $t_n'=t_0$ with $r_{i\to i+1}'=r_{n-i-1\to n-i}^{-1}$.
\end{definition}

\begin{definition}[Computational Inertia]
  \label{def:inertia}
  Computational Inertia in the $\slantedAleph$-calculus is the property that if, in a given reduction (Definition~\ref{def:reduction}), $r_{i\to i+1}=r$, then $r_{i+1\to i+2}$ cannot be $r^{-1}$.
  This would allow for a `futile cycle' in which no computational progress is made.
\end{definition}

\begin{theorem}[Non-Ambiguity]
  \label{thm:nonamb}
  A program is unambiguously/deterministically reversible if each variable appears exactly twice\footnote{In the examples in Section~\ref{sec:ex} the reader may notice this is violated. This is for programmer convenience, and must be resolved manually or by the compiler.} in a computational definition, and if and only if there exists no multiterm matching either (1) three-or-more computational rule patterns, or (2) two-or-more computational rule patterns and one-or-more halting patterns.
\end{theorem}

\section{Semantics}
\label{sec:sem}

As stated earlier, a computation consists of a multiterm $t$ that evolves reversibly in the context of a program environment, $\mathcal P$.
$\mathcal P$ is a set containing all the definitions making up the program, computational and halting.
The $\aleph$-calculus can be formulated without computational inertia (Definition~\ref{def:inertia}), but in the interest of brevity we give the inertial semantics here.
In the inertial semantics, each computational definition is assigned a pair of unique identifiers: $r$, corresponding to the left-to-right rule, and $r^{-1}$, corresponding to the right-to-left rule.
Each halting definition is assigned the identifier $\haltSym$, where for convenience $\haltSym^{-1}\equiv\haltSym$.
$\mathcal P$ is then a set of elements $r:\delta$ where $\delta$ is a definition and $r$ is the corresponding left-to-right rule identifier;
  for simplicity, $\mathcal P$ is extended with $r^{-1}:\delta^{-1}$ for each computational definition $\delta$ where $\delta^{-1}$ swaps the left and right sides of the head-rule.
Each multiterm $t=t_1\cdots t_n$ is then tagged as either halting, $[t_1\cdots t_n|\top]$ for initial multiterms and $[t_1\cdots t_n|\bot]$ for final multiterms, or `active', $(t_1\cdots t_n|r)$ where $r$ is the previously applied rule.
As well as tagging active multiterms with the previous rule application, it is important we identify halting tags because sub-multiterms must only be produced/consumed in a halting state or else non-determinism arises\footnote{The requirement of halting, combined with computational inertia, ensures each sub-multiterm takes on a unique state at production and before consumption.}.

At each computational step, we pick a non-final halting multiterm $t$ (either the `root' multiterm or a nested multiterm, \ruleName{sub--eval}) and match it against all the rules in $\mathcal P$.
The set $\mathcal M(t)$ is said set of matching rules, and is generated by rules \ruleName{unit,match}.
The rules for `matching' relate to `unification' and will be described later.
If the multiterm is initial-halting, and we find a match of the form $\haltSym:\delta\in\mathcal P$ then we can map $[t|\top]\to(t|\haltSym)$;
  if there aren't any halting definitions, then the computation enters an invalid state.
If the multiterm is active, then there are four valid cases:
  (1) The tagged multiterm is $(t|\haltSym)$ and $t$ matches \emph{only} one-or-more halting definitions $\haltSym:\delta$; we map $(t|\haltSym)\to[t|\bot]$.
  (2) The tagged multiterm is $(t|\haltSym)$ and $t$ matches one-or-more halting definitions and precisely \emph{one} computational definition $r:\delta$; if $r$ maps $t\ruleto{r} t'$, then we map $(t|\haltSym)\to(t'|r)$.
  (3) The tagged multiterm is $(t|r)$ and $t$ matches two computational definitions, $r^{-1}:\delta$ and $s:\varepsilon$, where $s$ may be $r$ but not $r^{-1}$; then, if $s$ maps $t\ruleto{s} t'$, we map $(t|r)\to(t'|s)$.
  (4) The tagged multiterm is $(t|r)$ and $t$ matches precisely \emph{one} computational definition, $r^{-1}:\delta$, and one-or-more halting definitions; we map $(t|r)\to[t|\bot]$.
These five cases above are given by rules \ruleName{step--halt,step--comp}.
Any other cases cause computation to enter an invalid state, either because there is no rule specified to continue the multiterm's evolution, or because there is an ambiguity in $\mathcal P$.
Note that each of the above tagged maps are reversible because whenever we consume $r$ it is by identifying exactly two possible rule identifiers, $\{r^{-1},s\}$, and using this information to consume the old tag $r$ and replace it with $s$.

\begin{listing}[tb]
  \fboxsep=0pt
  \fbox{%
    \begin{minipage}{\linewidth}
      \vspace{.5\baselineskip}%
      \begin{equation*}\begin{gathered}
\\[-1.5\baselineskip]
\begin{aligned}
\cfrac{\bigwedge_i s_i\to t_i}{s_1\cdots s_n \to t_1\cdots t_n}~
\ruleName{sub--eval} &&
\cfrac{\sigma \in \bnfPrim{sym}}{\sigma \unif\sigma \varnothing}~~
\cfrac{v \in \bnfPrim{var}}{t \unif v \{v\mapsto t\}}~~
\cfrac{\bigwedge_i t_i \unif{\tau_i} T_i\quad \bigcap_i T_i=\varnothing}%
      {[t_1\cdots t_n|\bot] \unif[1.75]{\tau_1\cdots\tau_n} \biguplus_i T_i}~
\ruleName{unif}
    \end{aligned} \\
    \begin{aligned}
\cfrac{}{\haltSym:\haltSym; \in \mathcal P}~
\ruleName{unit} &&
\cfrac{\haltSym:\haltSym\pi;\in\mathcal P \quad t\unif{\pi} V}%
      {\haltSym\in\mathcal M(t)}~~
\cfrac{r:(\lambda{=}\rho{:}\Sigma)\in\mathcal P \quad t\unif{\lambda} V}%
      {r\in\mathcal M(t)}~
\ruleName{match}
    \end{aligned} \\
    \begin{aligned}
\cfrac{\haltSym\in\mathcal M(t)}{[t|\top]\to(t|\haltSym)}~~
\cfrac{\mathcal M(t)=\{\haltSym\}}{(t|\haltSym)\to[t|\bot]}~~
\cfrac{\mathcal M(t)=\{r^{-1},\haltSym\}}{(t|r)\to[t|\bot]}~
\ruleName{step--halt}
    \end{aligned} \\
    \begin{aligned}
\cfrac{\mathcal M(t)=\{\haltSym,r\} \quad t\ruleto{r}t'}%
      {(t|\haltSym)\to(t'|r)}~~
\cfrac{\mathcal M(t)=\{r^{-1},s\} \quad t\ruleto{s}t'}%
      {(t|r)\to(t'|s)}~
\ruleName{step--comp}
    \end{aligned} \\
    \begin{aligned}
\cfrac{r:(\lambda{=}\rho{:}\Sigma)\in\mathcal P
        \quad \bigwedge_i t_i \unif{\lambda_i} T_i
        \quad \bigcap_i T_i=\varnothing}%
      {t_1\cdots t_n \ruleto{r}_{\mathit{init}}
        \langle r,\Sigma,\varnothing,\uplus_i T_i\rangle}~
\ruleName{comp--init}
    \end{aligned} \\
    \begin{aligned}
\cfrac{r:(\lambda{=}\rho{:}\Sigma)\in\mathcal P
        \quad \bigwedge_i t_i \unif{\rho_i} T_i
        \quad \bigcap_i T_i=\varnothing}%
      {\langle r,\varnothing,\Sigma,\uplus_i T_i\rangle
       \ruleto{r}_{\mathit{fin}} t_1\cdots t_n}~
\ruleName{comp--fin}
    \end{aligned} \\
    \begin{aligned}
\cfrac{[s|\top]\to [t|\bot]\quad [s]\unif\sigma S\quad [t]\unif\tau T}%
      {\langle r, \{\sigma{=}\tau{.}\}\cup\Sigma,
                  \Sigma',R\uplus S\rangle
       \ruleto{r}_{\mathit{sub}}
       \langle r,\Sigma,\{\sigma{=}\tau{.}\}\cup\Sigma',
                 R\uplus T\rangle}~
\ruleName{comp--sub$_\ell$}
    \end{aligned} \\
    \begin{aligned}
\cfrac{[s|\top]\to [t|\bot]\quad [s]\unif\sigma S\quad [t]\unif\tau T}%
      {\langle r, \{\tau{=}\sigma{.}\}\cup\Sigma,
                  \Sigma',R\uplus S\rangle
       \ruleto{r}_{\mathit{sub}}
       \langle r,\Sigma,\{\tau{=}\sigma{.}\}\cup\Sigma',
                 R\uplus T\rangle}~
\ruleName{comp--sub$_r$}
    \end{aligned} \\
    \begin{aligned}
\cfrac{s_1\cdots s_m\ruleto{r}_{\mathit{init}}^{\raisebox{-1ex}{}}%
            \ruleto{r}_{\mathit{sub}}^{%
            \kern-0.1ex\raisebox{-1ex}{$\scriptstyle\ast$}}
        \ruleto{r}_{\mathit{fin}}^{\raisebox{-1ex}{}} t_1\cdots t_n}%
      {s_1\cdots s_m\ruleto{r} t_1\cdots t_n}~
\ruleName{comp} &&
\cfrac{s\in\bnfPrim{term}^\ast}{s \to s}~~
\cfrac{s\to t\quad t\to u}{s\to u}~
\ruleName{closure}
\end{aligned}
\\[-0.5\baselineskip]
    \end{gathered}\end{equation*}%
    \vspace{.5\baselineskip}%
  \end{minipage}}%

  \caption{%
    The semantics of the (inertial) $\aleph$-calculus.
    Note that $\langle r,\Sigma\,\Sigma',S\rangle$ represents an intermediate computational state, where $r$ is the rule identifier, $\Sigma$ is a set sub-rules yet to be applied, $\Sigma'$ is a set of sub-rules that have been applied, and $S$ is a set of variable bindings.}
  \label{lst:sem}
\end{listing}

In the above, we relied on the notion of a rule $r:(\lambda{=}\rho{:}\Sigma)$ inducing the mapping $t\ruleto{r} t'$.
The semantics, given by rules \ruleName{comp,--init,--sub$_\ell$,--sub$_r$,--fin}, are as follows:
The pattern $\lambda$ is unified against $t$: here, unification means that $t$ is reversibly consumed by comparison with the pattern $\lambda$, in the process producing a variable mapping.
Then we pick (without replacement) a sub-rule from $\Sigma$ where one side consists only of variables in our current mapping.
We apply this sub-rule: the variables are substituted into the pattern, the resulting multiterm $t$ is instantiated as $[s|\top]$ and evolved to $[s'|\bot]$, and $s'$ is then matched against and consumed by the other pattern of the sub-rule.
This can fail, leading to an invalid state, if the new multiterm doesn't evolve to $[s'|\bot]$ or if $s'$ doesn't unify with the final pattern.
This process is repeated until our current variable mapping can be substituted into $\rho$ to yield $t'$, completing the mapping of $t\to t'$.

Unification, rules \ruleName{unif}, is simply recursive pattern-matching.
If a term $t$ unifies against a pattern $\pi$ with variable bindings $V$, we write $t \unif{\pi} V$.
The pattern $\sigma\in\bnfPrim{sym}$ only unifies with the term $\sigma$.
The pattern $v\in\bnfPrim{var}$ unifies with any term $t$ with bindings $\{v\mapsto t\}$.
The pattern $(\pi_1\pi_2\cdots \pi_n)$ unifies with a multiterm $[t_1t_2\cdots t_n|\bot]$ if and only if each $\pi_i$ unifies with $t_i$ with bindings $V_i$, and the $V_i$ are disjoint (i.e.\ the $\pi_i$ don't share any variables);
  notice that it only unifies with a final-halting state.

These semantics are summarized in Listing~\ref{lst:sem}.
Complete computation is achieved via the rules \ruleName{closure}.
An immediate concern is that the application of sub-rules and the evolution of sub-multiterms is non-deterministic/asynchronous.
In fact this is a feature, and allows for automatic parallelisation of independent subcomputations in the $\aleph$-calculus.
A necessary condition is confluence, which is satisfied by the $\aleph$-calculus (Theorem~\ref{thm:confl}).
Finally it is important for the usefulness of the $\aleph$-calculus that it is reversible-Turing complete (Theorem~\ref{thm:rtc}).

\begin{theorem}
  \label{thm:confl}
  The semantics of the \slantedAleph-calculus are confluent, in the sense that the final result is independent of evaluation order (and possible parallel evaluation).
\end{theorem}

\begin{theorem}
  \label{thm:rtc}
  The \slantedAleph-calculus is reversible-Turing complete, in the sense that it can reversibly simulate (without additional garbage) the reversible Turing Machine defined by Bennett~\cite{bennett-tm} and vice-versa.
\end{theorem}

\section{Discussion \& Future Work}
\label{sec:concl}

We have introduced a novel model of reversible computing, the $\aleph$-calculus, that is declarative and has a TRS semantics.
We proved (see extended paper) that the calculus is non-ambiguous, reversible-Turing complete, and that its semantics are confluent.
An interpreter has also been written and is available online\footnote{\url{https://github.com/hannah-earley/alethe-repl}}.
It may also be extended to support concurrency, with interesting consequences for determinism and causal-consistency.
The concurrent and non-inertial variant of the calculus, introduced in the extended paper, gives an alternate positioning of the model in the context of molecular programming:
  another form of unconventional computing, in which the interactions of specially prepared molecules simulate computation (see Zhang and Seelig~\cite{dsd-review} for a review of one such approach).

\section*{Acknowledgements}

  The author would like to acknowledge the invaluable help and support of her PhD supervisor, Gos Micklem.
  This work was supported by the Engineering and Physical Sciences Research Council, project reference 1781682.

\bibliography{references}{}
\bibliographystyle{splncs04}
\end{document}